\documentclass[conference]{IEEEtran}
\IEEEoverridecommandlockouts
\usepackage{cite}
\usepackage{amsmath,amssymb,amsfonts}
\usepackage{algorithm}
\usepackage{algorithmic}
\usepackage{graphicx}
\usepackage{textcomp}
\usepackage{xcolor}
\usepackage{caption}
\usepackage{subcaption}
\usepackage{comment}
\usepackage{multirow}
\usepackage{diagbox}
\usepackage{balance}

\usepackage{amsthm}

\usepackage{enumitem}

\usepackage{color, colortbl}
\definecolor{Gray}{gray}{0.9}
\definecolor{LightCyan}{rgb}{0.88,1,1}

\newcolumntype{a}{>{\columncolor{Gray}}c}
\newcolumntype{b}{>{\columncolor{LightCyan}}c}
\def\BibTeX{{\rm B\kern-.05em{\sc i\kern-.025em b}\kern-.08em
    T\kern-.1667em\lower.7ex\hbox{E}\kern-.125emX}}

\usepackage{pifont}

\begin{document}

\title{Job Scheduling in High Performance Computing}

\author{\IEEEauthorblockN{Yuping Fan}
\IEEEauthorblockA{\textit{Illinois Institute of Technology}}
}

\maketitle

\begin{abstract}
The ever-growing processing power of supercomputers in recent decades enables us to explore increasing complex scientific problems. Effective scheduling these jobs is crucial for individual job performance and system efficiency. The traditional job schedulers in high performance computing (HPC) are simple and concentrate on improving CPU utilization. The emergence of new hardware resources and novel hardware structure impose severe challenges on traditional schedulers. The increasing diverse workloads, including compute-intensive and data- intensive applications, require more efficient schedulers. Even worse, the above two factors interplay with each other, which makes scheduling problem even more challenging. In recent years, many research has discussed new scheduling methods to combat the problems brought by rapid system changes. In this research study, we have investigated challenges faced by HPC scheduling and state-of-art scheduling methods to overcome these challenges. Furthermore, we propose an intelligent scheduling framework to alleviate the problems encountered in modern job scheduling.
\end{abstract}


\section{Introduction}
Supercomputers are experiencing substantial changes. The new hardware resources, such as burst buffer, are emerging. As new resources are incorporated into the system, schedulers need to consider new resources in decision making process. At the same time, the hardware structures are changing rapidly. For example, the shared network structures, such as Dragonfly and fat tree, are widely adopted in new generation supercomputers. The changes in hardware structures complicate the scheduling problem, especially when the hardware becomes sharable. Sharing resources is one of the most effective methods to improve resource utilization, but it also brings many issues, such as resource fairness and resource contention. Without scheduling shared resource properly, system performance is suffered from resource contention and fairness will be hurt too. In order to fully utilize resources, schedulers are required to keep up with the changes in hardware.
The substantial improvement in hardware drives users to submit more complex problems. For example, supercomputers having GPU attract deep learning projects, which require a massive number of threads to tolerate latencies. Supercomputers with burst buffer enable data-intensive applications, such as scientific simulations, to execute. Schedulers are responsible for monitoring and cooperating resources in a system to support rapid service to users. Without central control from schedulers, jobs could easily fail and the failure could propagate to the whole system.
The hardware updates cause users to change their behavior to adapt to these changes. For example, the runtime variability brought by sharing network makes users more conservative in job runtime prediction \cite{Li1}. As supercomputers become more powerful, users tend to submit jobs more frequently. Schedulers need to detect the changes in user behavior and adjust scheduling policies in order to maintain good performance.
In light of these challenges, studies in modern schedulers develop various methods to improve system efficiency and individual job performance. These goals of the methods can be roughly divided into groups: fairness, resource utilization, job performance. However, these goals are often conflicting. In this chapter, I review modern scheduling methods focusing on the applicability of the schedulers. In addition, an intelligent HPC job scheduling framework is proposed to address these issues.
The remainder of this chapter is organized as follows. I first review the existing HPC scheduling algorithms. I then introduce the proposed intelligent HPC job scheduling framework. Finally, I conclude this chapter.

\section{HPC Scheduling Algorithms}\label{HPC Scheduling Algorithms}
\subsection{Scheduling Algorithms on Production Systems}
Job scheduling in HPC is responsible for ordering jobs in waiting queue and allocating jobs to resources according to site policies and resource availability. In HPC, the well-known job schedulers include Slurm, Moab/TORQUE, PBS, and Cobalt. Similar to HPC job schedulers, cluster schedulers, such as Apollo, Mesos, Omega, and YARN, basically play the same role in a system. The main difference between HPC schedulers and cluster schedulers is that the infrastructures they served are different. HPC facilities are designed to serve big scientific and engineering applications that are impossible to run on other systems, while commercial clusters are inclined to serve big data applications. The scheduling algorithms used on both HPC systems and clusters are simple. The most widely adopted scheduling policies used on production HPC systems is FCFS (first-come first-served), which sorts waiting jobs in the order of their arrival times. For some leadership computing facilities, their main goals are enabling large jobs to run. Hence, jobs consuming more system resources have higher priorities to execute. For example, ALCF (Argonne Leadership Computing Facility) adopts a utility-based scheduling policy, named WFP, which favors large and old jobs in the queue \cite{Fan2}. Backfilling is a common strategy used in production HPC scheduling in order to enhance system utilization \cite{Tsafrir2007, Tsafrir2006}. The most widely used backfilling strategies are EASY backfilling and conservative backfilling. EASY backfilling allows waiting jobs to skip ahead under the condition that they do not delay the job at the head of the queue. Conservative backfilling has the stricter condition that jobs can be backfilled only if they do not delay the preceding jobs. Because clusters serve a different purpose than HPC systems, cluster schedulers concentrate on providing satisfactory service to all users by fair sharing the resources in a system.

\subsection{Hierarchical Scheduling Framework vs. Distributed Scheduling}
The ever-increasing HPC system scale poses a serious challenge to modern HPC scheduling. The current central scheduling model cannot keep up with the challenge of the increasing complexity of resource constraints. The emergence of hierarchical scheduling framework expands the traditional schedulers’ view beyond the single dimension of nodes. Flux is a good representative example of hierarchical schedulers \cite{Ahn}. In Flux, each job is an instance of the scheduling framework, which can launch many small and high-throughput sub-jobs. Therefore, it combats the issue of scalability that exists in many modern HPC schedulers. Resource management in Flux is operated at large granularity and it can move resources between child jobs. Because of the recursive feature of the hierarchical scheduling framework, it can be extended to schedule emerging resources with small changes.
Besides the hierarchical scheduling framework, the distributed scheduler is another scheduling framework to overcome the problem of scalability \cite{Dogar}. In a distributed scheduler, each job waiting in the queue are assigned to a distributed scheduler, which has their own resources. The resource exchanges are done by communication between distributed schedulers. The advantage of distributed schedulers is their scalability, because they require even less computation and memory than hierarchical scheduler, but the downside is the inefficiency of using system resources due to the isolation of system resources.

\subsection{Multi-Resource Scheduling}
Multi-resource scheduling is a research topic in HPC scheduling that raises more attention in recent years. This is because increasing more resources are incorporated into the next-generation HPC systems \cite{Hung, Qiao1, Qiao2, Fan3, Fan7}. A large body of multi-resource scheduling focuses on power and compute resource scheduling \cite{Zhou, xu01, Topper, Patki, Mammela, Guzek}. This requires the trade-off between power and performance in decision making. For example, Wallace et al. addressed the power-aware scheduling problem in HPC by optimizing compute node utilization with the power constraint \cite{Wallace}. This solution prefers to compute node utilization over power constraint.
In multi-resource cluster scheduling, fairness is more important than other factors. Dominant Resource Fairness (DRF) is a strategy to achieve fair allocation of various resources to users \cite{Ghodsi}. Although DRF maximizes resource fairness in a system and therefore obtains user satisfaction, much recent research found that fair sharing and high utilization are conflicting goals and aggressively using fair sharing have a negative effect on resource utilization. In order to address this challenge brought by fair sharing, some studies make tradeoffs between fairness and utilization. For example, Grandl et al. leveraged a multi-dimensional bin packing algorithm to improve resource utilization and then used a knob to balance resource utilization and fairness \cite{Grandl2014}. The advantage of using the bin packing algorithm is its speed, but it is also a greedy algorithm which allocates jobs in a one-by-one manner based on isolated job information. In an HPC system, a scheduler has more time to make scheduling decision but are required to make the best use of various resources compared to cluster scheduling. Therefore, multi-resource HPC scheduling demands more complicated scheduling methods. Optimization methods, especially multi-objective optimization methods, are leveraged to achieve better system performance in HPC scheduling \cite{Fan3, Fan4, Fan7}.

\subsection{Energy-, Power-, Cooling-Aware Scheduling}
Energy consumption in HPC and datacenters raise great attention in recent years. As HPC systems and datacenters become increasingly more powerful, one side effect is that the generated power cannot keep up with their consumption rate. For example, in 2014, data centers in the U.S. consumed 1.8\% of the total electricity consumption (70 billion kWh of energy). Because energy expense is becoming an increasingly dominant portion of the operation cost in HPC and data centers, data centers and HPC systems attempt to reduce energy consumption without hurting performance through more effective job scheduling strategies and more energy efficient hardware. The Dynamic Voltage and Frequency Scaling (DVFS) is technique widely studied in the literature \cite{Lee}, which adjusts power and speed settings on a computing device’s various processors, controller chips and peripheral devices to optimize resource allotment for tasks and maximize power saving when those resources are not needed. In addition, shutting down unused hardware is another effective way of lowering power consumption. However, this method comes at the cost of waste system resources. The time of turning on hardware resources can cause a delay in latency-critical jobs and tasks. Although, this approach is widely studied but it rarely used in real systems. In recent decades, usage of renewable and sustainable energy source to replace traditional energy, such as fossil fuels, is a hot topic. Renewable energy is energy that is collected from renewable resources, which are naturally replenished on a human timescale, such as sunlight, wind, rain, tides, waves, and geothermal heat. Although there is a great promotion in using renewable energy, renewable energy only contributed 19.3\% to humans' global energy consumption in 2017. Therefore, in recent years, HPC and data center facilities try to cut their traditional energy consumptions \cite{Pahlevan, Kong, Devabhaktuni, Garg, Goiri, Kliazovich}. Therefore, there is a research area focusing on maximizing resource consumptions from renewable energy so as to reduce energy consumption from traditional energy sources. Besides energy costs are fluctuating, the energy price is often low at night when the energy consumption is lowest in general, while the price is high during the daytime. In order to save energy costs, scheduling methods are developed to use more energy when the energy costs are low and decreasing energy consumption when the energy price is high \cite{xu01}.

\subsection{HPC Failures and Scheduling}
As the size of HPC systems increases drastically in recent years, failures in HPC systems increases accordingly. One of the most important factors of increasing application failures is the increase in application sizes. Any hardware failure can cause application failure \cite{Tang, Vinay, Kumar}. A large application using many resources in a system such as compute resource, the network resource is more likely to meet hardware failure than a small application \cite{Webb}. Aging is another factor that causes failures of HPC applications. For example, Zimmer et al. analyzed the relationship between GPU aging with the reliability of HPC jobs on Titan \cite{Zimmer}. The analysis presents that large applications (use more than 20\% of machine resources) encounter a higher level of application failures. Therefore, in their work, they replaced 50\% of aged GPUs and employed techniques to use low- failure GPUs to run large jobs through targeted resource allocation. By deploying the techniques in a real HPC system, Titan, they demonstrated the positive impacts of age-aware resource allocation policy. To tolerate failures in the HPC environment, applications can implement checkpoint technique, which pauses the application and then copies all the required data from the memory to reliable storage and then continues the execution of the application. In case of failure, the application could restart from the latest checkpoint from the stable storage and execute from there. Therefore, the checkpoint technique avoids the trouble of starting from scratch. Checkpointing uses many system resources, such as memory, network, and storage system. Therefore, the best checkpoint frequency raises much research interests in recent years \cite{Gomez, Fan8}. Frequent checkpointing introduces too many overheads to an application; however, if this application fails, it can be restarted from a very recent point. There are two checkpointing levels: global checkpointing and local checkpointing. Local checkpointing is light-weighted, which only copies memory within a node or a group, while global checkpointing copies the whole application’s memory and it requires global consistency. However, local checkpointing can only recover local hardware failures. If the cascading failures happened, the application has to roll back to the latest global checkpoint. Because checkpointing consumes resource times, it needs to be considered into scheduling. For example, when users make their runtime estimates on their jobs, if they use checkpoints, they need to take the time consumed by checkpoints into consideration. In addition, checkpointing operation consumes system resources, such as memory for local checkpointing and network, I/O and storage resources. When users reserve their resources at job submission, they need to estimate the additional resources consumed by checkpointing operations.

\subsection{Backfilling and User Runtime Estimates}
Backfilling is a common strategy used by production HPC facilities. The widely known backfilling strategies are EASY backfilling and conservative backfilling. EASY backfilling is the easiest to implement and it produces good scheduling results on production systems, and therefore it is the most widely used strategy. Besides these two backfilling strategies, there are some other strategies proposed to improve the performance of backfilling. One of the most effective methods is to improve user runtime estimates. User runtime estimate is the upper bound of a job’s runtime. If a job needs more than this upper bound, this job will be killed by the scheduler. This phenomenon is called underestimation. If a job finished before it reaches this upper bound, this user overestimates this job. The
accuracy of user runtime estimate (defined as $\frac{Job\_Actual\_Runtime}{Job\_Runtime\_Estimate}$) is very 
important factors in scheduling, because all scheduling decisions are made based on user provided runtime estimates. However, user provided runtime estimates are proved to be very inaccurate. Based on the previous studies, the average accuracies of job runtime estimates on many production systems are less than 60\% \cite{Fan1}. Scheduling decisions made by on these inaccurate runtime estimates cause many scheduling performance problems, such as low resource utilization, backfilling and job priority issues. Therefore, it is crucial to provide more accurate runtime estimates. There are several methods in literature to improve user runtime estimates. With the widely used of machine learning algorithms in various scientific fields, there are several attempts to leverage machine learning algorithms to predict job runtimes \cite{Fan1,Gaussier}. The basic idea of these machine learning approaches is to extract features and make job runtime predictions from user inputs, such as job runtime estimates and job size, and historical job information, such as the job runtimes from the same user and project. For example, Gaussier et al. leveraged an online linear regression model to predict job runtime \cite{Gaussier}. Fan et al. extended Tobit model to balance runtime prediction accuracy and underestimation rate \cite{Fan1}. Improving runtime estimates is one way to enhance scheduling performance, optimizing backfilling is another effective way to improve system resource utilization. The traditional backfilling strategies (EASY backfilling and conservative backfilling) picks jobs to backfilled from the front of the queue. Once they find a job that fit the hole in the schedule, they will backfill this job immediately. However, this selection approach may miss the best matching jobs. In addition, to avoid jobs to be killed by systems due to underestimation, methods are used to correct prediction adaptively and this approach allows users to provide more accurate runtime estimates.

\subsection{Scheduling Moldable and Malleable Jobs}
Based on who decides the number of nodes and when it is decided, HPC jobs can be classified into four categories: rigid, evolving, moldable, and malleable. For rigid and evolving jobs, users decide how many nodes to be used. For rigid jobs, the decisions are made at submission, while for evolving jobs, users can change their node requests during execution. For moldable and malleable jobs, users specify a range of nodes a job can be run on and the scheduler decides how many nodes to be used. For moldable jobs, the scheduler makes decisions at submission. Malleable jobs are those which can dynamically shrink or expand resources on which they are executing at runtime. Executing moldable and malleable jobs can potentially improve system utilization and reduce average response time \cite{Fan9}. Executing those jobs are challenging for HPC systems and schedulers. At present, most HPC facilities do not support moldable and malleable jobs. First, the nature of HPC applications makes it difficult to change job size during execution. Most HPC applications have intensive communication between nodes, which makes dynamically changing job size very difficult \cite{Qiao1, Qiao2}. Second, this requires HPC schedulers to be adaptive and monitor system status. Therefore, enabling malleable jobs on HPC systems demands HPC schedulers to do more jobs in a very short time. Despite the forehead mentioned challenges, there are studies attempting to scheduling malleable jobs in order to improve system performance \cite{Gupta, Sadykov}.

\subsection{Workflow Scheduling}
Data-intensive data analysis applications often utilize a workflow that contains tens or even hundreds of tasks. Jobs are made of stages, such as map or reduce, lined by data dependencies. When a task has all the required input data ready, it will be allocated resources to execute. The input data is stored in file systems and is divided into multiple chunks, each of which is typically replicated three times across different machines. Therefore, executing these data-dependent tasks need to follow the strict order feed by users. In addition, choosing the location for each task is critical for efficiency in executing jobs. Schedulers prefer to execute tasks on the machines that have a copy of the input data, because local access of input data could save the time on transfer input on network. It is also beneficial to the whole systems, because it reduces the amount of data moved in a global network \cite{Jalaparti}. Therefore, there are some studies concentrating on improving data locality in scheduling \cite{Ahmad, Caniou, Masdari, Estrada}. For example, Quincy attempts to balance between latency and data locality of all runnable tasks \cite{Isard}. The data locality of MapReduce jobs can be improved by scheduling both map and reduce tasks of one job on the same rack. Corral achieves better data locality by coupling the placement of data and compute nodes. Data analysis applications are often delay-sensitive, which means it is crucial for meeting the deadlines of these workflows. Workflow scheduling problem is known to be NP-complete in general. Scheduling algorithms often utilize heuristics and optimization techniques to try to obtain a near optimal scheduling decision. Hadoop is a popular map-reduce implementation deal with independent map-reduce tasks. To meet deadline satisfaction, a large body of studies concentrate on giving higher priorities to time-sensitive tasks and delay other tasks or jobs in a system so as to reduce deadline violations of time-sensitive tasks. Delay scheduling meet the deadline via another approach, which allows time-sensitive tasks to wait for a certain amount of time and this increases the chance of finding a better allocation for time-sensitive tasks that can store data locally \cite{Zaharia}. Apache Oozie is a workflow scheduler for Hadoop jobs, which presents workflow tasks as Directed Acyclic Graphs (DAGs).Oozie combines multiple jobs sequentially into one logical unit of work, and gives the provision to execute tasks which are scheduled to run periodically.

\section{Overview of Proposed Job Scheduling Framework}
The challenges faced in today’s job scheduling in HPC and data centers demand more intelligent schedulers to make smarter scheduling decisions. Therefore, I extend the modern scheduling framework for job scheduling in HPC, which comprises of a resource manager, a job manager, a scheduling decision maker, a system performance monitor, and a job performance monitor. The functions of these models are explained as follows:
\begin{enumerate}
    \item Resource manager: Unlike the traditional resource manager which only manages nodes, the next-generation resource managers are responsible to monitor the status of various schedulable resources in the HPC system, allocate resources to jobs, retrieve resource when job finished or failed, and report abnormal resource behaviors.
    \item Job manager: Upon job submission, a job manager records the job’s basic information (such as user name, project name) and its resource requirement (such as node requirement, the maximum time to run the job, and memory requirement). The job manager informs scheduling decision maker about the basic information of the incoming job and the scheduling decision maker orders jobs in the waiting queue based on user input and current status in the waiting queue. In addition, the job manager is responsible to monitor the job status changes and update the job status based on the information provided by the resource manager, scheduling decision maker and OS system.
    \item Scheduling decision maker: To meet the challenges in modern job scheduling in HPC, the next-generation decision maker needs to make smart scheduling decision based on the flexible requirement from the system administrators. Different HPC systems and data centers have their unique goals. For example, some systems aim to provide fast response time to time-sensitive applications, and the decision maker is supposed to adopt the scheduling algorithms that give high priorities to the time-sensitive applications and reserve a portion of system resources to meet the burst of the time-sensitive applications. If an HPC system concentrates on running big jobs, the scheduler needs to focus on assign big jobs to the allocation with minimal interference from other jobs. A scheduler used in a production system is supposed to be flexible to plugin other scheduling policies.
    \item System performance monitor: If a system focuses on optimizing its resource utilization, the system performance monitor is needed to monitor the current system status and record and analyze the system past performance. The purpose of monitoring and analysis is to alert system administrators and correct scheduling decisions if the system performance does not reach the expectation \cite{Fan5, Fan6}.
    \item Job performance monitor: The job performance monitor record and report abnormal job behaviors, such as abnormal exit from the system, and long job wait time and long job running time. The analysis can also be conducted on user or project based, so the system administrators can find what kind of user behavior or application source code can cause the degradation of job performance.
\end{enumerate}

In summary, an intelligent job scheduler is the trend in the future. The main difference of the intelligent job scheduler and the traditional job scheduler is that the intelligent job scheduler is capable of monitoring job and system status and therefore provide feedback to system administrators and schedulers itself to make adjustment accordingly.

\section{Conclusion}
Job scheduling in HPC systems and data centers is one of the active research fields which plays a crucial role in effective utilization HPC and data center resources and efficient execution of jobs. In this chapter, I reviewed the challenges faced by HPC job scheduling and the approaches adopted by schedulers to alleviate these problems. From the literature review, I found that the current HPC job scheduling framework is not smart to address various challenges. Therefore, I propose an intelligent HPC job scheduling framework to monitor the abnormal behaviors and performance in an HPC system and improve system and job performance dynamically.

\balance
\bibliographystyle{unsrt}
\bibliography{ipdps1.bib}

\end{document}